\documentclass[aps,prd,showpacs,preprintnumbers,nofootinbib,floatfix,floats,groupedaddress,twocolumn]{revtex4}

\usepackage{bm}
\usepackage{latexsym}
\usepackage{dcolumn}
\usepackage{amsmath,amsfonts,amssymb}
\usepackage{graphicx,epsfig}
\usepackage{color}
\usepackage[active]{srcltx}
\def\nn{\nonumber}

\def\l{\left}
\def\r{\right}
\def\DM{\mathrm{d}}

\reversemarginpar

\def \mQ {\Psi}
\def \gm {{\overset{\star}{g}}}
\def \boxm {{\overset{\star}{\square}}}
\def \gdummy {{\overset{\textcolor{white}{\star}}{g}}}
\def \lp {L_0}
\def \lpp {\ell_{\star}}
\def \mA {\mathcal A[\sigma;\lp]}

\begin{document}

\title{Minimal Length and Small Scale Structure of Spacetime}

 \author{Dawood Kothawala}
 \email{dawood@physics.iitm.ac.in}
 \affiliation{Department of Physics, Indian Institute of Technology Madras, Chennai 600 036}

\date{\today}
\begin{abstract}
\noindent Many generic arguments support the existence of a minimum spacetime interval $\lp$. Such a ``zero-point" length can be naturally introduced in a locally Lorentz invariant manner via Synge's world function bi-scalar $\Omega(p,P)$ which measures squared geodesic interval between spacetime events $p$ and $P$. I show that there exists a {\it non-local} deformation of spacetime geometry given by a {\it disformal} coupling of metric to the bi-scalar $\Omega(p,P)$, which yields a geodesic interval of $\lp$ in the limit $p \rightarrow P$. Locality is recovered when $\Omega(p,P) \gg \lp^2/2$. I discuss several conceptual implications of the resultant small-scale structure of spacetime for QFT propagators as well as spacetime singularities.
\end{abstract}

\maketitle
\vskip 0.5 in
\noindent
\maketitle
\section{Introduction} \label{sec:intro} 
Existence of a minimal length scale is one of the most generic implications of attempts to combine the fundamental principles of quantum field theory (QFT) and General Relativity (GR) \cite{sabine-lrr}. This length scale is often believed to be of the order of Planck length, i.e., $L_0=O(1) \lpp$ where $\lpp=(G\hbar/c^3)^{1/2}=10^{-33}$ cm. 
\footnote{One exception is the Hoyle-Narlikar action-at-a-distance theory, in which the relevant QFT cut-off comes from ``response of the universe" \cite{HN-rmp}. I thank Prof. Narlikar for pointing this out to me.}
, and serve as a universal regulator for divergences in QFT and the singularities in classical solutions of GR \cite{deser-1957}. Various attempts have been made in the past to quantify this idea at the level of effective field theory. In particular, Bryce DeWitt in 1981 proposed an effective action for quantum gravity, which was motivated by his earlier result \cite{dewitt-1964} in which he studied the Bethe-Salpeter amplitudes for graviton exchange between scalar particles in the ladder approximation, and found that resummation of relevant diagrams implies (in the limit of large momentum transfer) that the singularity of the effective gravitational interaction is shifted away from the light cones, onto a spacelike hyperboloid of size $L_0$. Specifically, the structure of the propagator acquires a modification of the form 
$2 G_{\infty}(p,P) \equiv \l[\Omega(p,P)\r]^{-1} \rightarrow \l[\Omega(p,P) - 2 L_0^2\r]^{-1}$, where 
\begin{eqnarray}
\Omega(p,P) = \frac{1}{2} \l( \lambda(P) - \lambda(p) \r) \int \limits_{\lambda(p)}^{\lambda(P)} \l[ g_{ab} t^a t^b \r] \l(x(\lambda)\r) \DM \lambda
\end{eqnarray}
is the Synge world function bi-scalar \cite{poisson-lrr} and $t^a$ is tangent vector of the geodesic connecting event $P$ to $p$, with affine parameter $\lambda$. An independent line of work, based on quantum conformal fluctuations of the spacetime metric, was also shown to lead to similar result (see, for e.g., \cite{paddy-1985}). Yet another analysis leading to the same result, based on path-integral over gravitational field $h_{ab}=g_{ab}-\eta_{ab}$, was presented in \cite{ohanian-1999}. (Henceforth, for clarity, I will work with the geodesic distance $\sigma^2(p,P)=2 \Omega(p,P)$, and often employ coordinates $x=x(p)$, $X=X(P)$ etc. to represent spacetime events. I will also assume that $p \in \mathcal N(P)$, the normal convex neighbourhood of $P$. All differentiations in the manuscript are w.r.t. $x$.) 

The main implication of these older results can be captured in the following geometric, and {\it locally Lorentz invariant}, statement: if $\sigma(p,P|g_{ab})$ denotes the geodesic distance between the spacetime points $p$ and $P$ in the background metric $g_{ab}$, then quantum gravitational fluctuations lead to 
\begin{eqnarray}
\l \langle \sigma^2(p,P \vert (g \circ h)_{ab}) \r \rangle = \sigma^2(p,P \vert g_{ab}) 
+ \epsilon L_0^2
\label{eq:zpl-general}
\end{eqnarray}
where $``h_{\bullet \bullet \ldots}"$ represents {\it all} (scalar, tensor \ldots) possible quantum fluctuations of the background metric $g_{ab}$, and $(g \circ h)_{ab}$ symbolically represents the deformation of $g_{ab}$ produced by these fluctuations. The brackets $``\langle ... \rangle"$ represent a suitable path integral average over the fluctuations
\footnote{The factor $\epsilon=\pm 1$ represents an ambiguity in precisely how $L_0^2$ gets added to the spacetime intervals. Although relevant for singularity structure of the propagator, I will absorb this factor into $L_0^2$; it can always be restored by replacing $L_0^2 \rightarrow \epsilon L_0^2$.}
. A complete framework of quantum gravity is, of course, expected to specify both the precise form of the fluctuations and the prescription for averaging over them. In this context, it is worth mentioning that a particularly elegant connection exists between Eq.~(\ref{eq:zpl-general}) and the so-called ``duality" transformation of (infinitesimal) relativistic point particle action: $\DM s \rightarrow \DM s + \lp^2/\DM s$ ($\DM s$ is the arc-length), which essentially makes the path amplitude invariant under the inversion $\DM s \rightarrow \lp^2/\DM s$. This was pointed out by Padmanabhan in \cite{paddy-1997-1998}, and it's connection with the notion of T-duality in string theory was explored in \cite{fontanini-2006}; the latter was again shown to yield result consistent with Eq.~(\ref{eq:zpl-general}). 

Given it's relevance, it is therefore of significance to ask whether, given a spacetime metric $g_{ab}$, there exists some deformation of it which directly leads to the result in Eq.~(\ref{eq:zpl-general}). That is, whether one can find a metric $\gm_{ab}$ such that 
\begin{eqnarray}
\sigma^2(p,q \vert \gm_{ab}) \; = \; \sigma^2(p,q \vert g_{ab}) + L_0^2 
\end{eqnarray}
The idea that existence of a minimal length might require modification of geometry is not new \cite{march-1936}, although a precise, mathematical characterization of the modification has not been attempted before. In this Letter, I show that not only does there exist an {``effective spacetime metric"} $\gm_{ab}$ which yields the desired result, but it turns out to have an unexpectedly simple form in terms of a {\it disformal} coupling of the spacetime metric to the bi-scalar ${\mA }=1+\lp^2/\sigma^2(p,q)$. Several conceptual implications of this result are discussed.

The main results of this Letter are contained in Eqs.~(\ref{eq:key1}), (\ref{eq:key2}), (\ref{eq:boxss}) \& (\ref{eq:leading-sing}).

\section{Constructing the modified metric} \label{sec:2}
It is easiest to begin with flat spacetime, since this is the context in which the results quoted above were first derived. In flat spacetime, the propagator of a massive scalar field gets modified to
\begin{eqnarray*}
G(x,X) = \left(\frac{4 m^2}{- \sigma_{\eta}^2 - \lp^2}\right)^{1/2} H^{(2)}_{1}\left[m \left( - \sigma_{\eta}^2 - \lp^2 \right)^{1/2}\right]
\end{eqnarray*}
where $\sigma_{\eta}^2 = \eta_{ab} (x^a-X^a) (x^b-X^b)$. I take this as my starting point, and ask if it is possible to deform/re-parametrize the standard flat Minkowski metric $\eta_{ab}$ such that the above propagator naturally arises as the kernel of the modified d'Alembartian operator. Then, {\it if} this modification is expressible in a covariant form, generalization to curved spacetime will be immediate. In fact, for flat spacetime, one can achieve the desired modification by introducing a ``hole" of size $\lp$ around the event $X$. This requires a diffeomorphism which is {\it singular} at $X$, and yields the following modified metric (see Appendix \ref{app:deriv-mod-metric} for details)
\begin{eqnarray}
\overset{\star}{\eta}_{a b}(x; X) &=& \l( 1 + \lp^2/\sigma_{\eta}^2 \r) \eta_{a b} - \Xi \; 
(x_a-X_a) (x_b-X_b)
\nn \\
\Xi[\sigma_{\eta}; \lp] &=& \l({1}/{\sigma_{\eta}^2}\r) \l(\lp^2/\sigma_{\eta}^2\r) 
\left( 1 +\l[{1 + \lp^2/\sigma_{\eta}^2}\r]^{-1} \right)
\label{eq:dm-metric}
\end{eqnarray}
The argument $X$ in $\overset{\star}{\eta}_{a b}(x; X)$ is important; it is a reminder that the modified metric depends on two-points, the field point $x$ and the base point $X$ around which the modification is being sought; this metric may then be interpreted as an ``effective metric" near $X$. 

To generalize to curved spacetime, note that: $\nabla^{(x)}_a \sigma_{\eta}^2 = 2 \eta_{a b} (x^b-X^b)$ in flat spacetime. This allows us to rewrite (\ref{eq:dm-metric}) as [replacing $\eta_{ab} \rightarrow g_{ab}$, $\sigma_\eta \rightarrow \sigma_{\bm g}$]:
\begin{eqnarray}
\gm_{a b}(x; \sigma_{\bm g}^2) &=& \l( 1 + \lp^2/\sigma_{\bm g}^2 \r) g_{a b}(x)
- \frac{\lp^2}{4}  \l( \frac{2 + \lp^2/\sigma_{\bm g}^2}{1 + \lp^2/\sigma_{\bm g}^2 } \r)
\nn \\
&\times& 
\partial_a \l(\ln |\sigma_{\bm g}^2|\r) \; \partial_b \l(\ln |\sigma_{\bm g}^2|\r)
\label{eq:pid-metric-final-presc}
\end{eqnarray}
This is the desired covariant form of Eq.~(\ref{eq:dm-metric}), and I will now show that $\gm_{ab}$ can be recast into a remarkably simple form, which is essentially a {\it disformal} transform \cite{disformal} of the original metric involving the bi-scalar $\Omega(x,X)$.

To do this, recall that the affinely parametrized tangent to the geodesic from $p$ to $P$ at $p$ (and pointing away from $P$) is given by $t_a = \l( \partial_a \sigma^2 \r)/ 2 \sqrt{\epsilon \sigma^2}$, where $t^2=\epsilon=\pm 1$. Using this, we finally obtain
\begin{eqnarray}
\gm_{ab}(p;P) &=& \mathcal{A} g_{ab}(p) - \epsilon \l( \mathcal{A}  - \mathcal{A}^{-1} \r) t_a t_b
\nn \\
\gm {}^{ab}(p;P) &=& \mathcal{A}^{-1} \; g^{ab}(p) + 
\epsilon \l( \mathcal{A}  - \mathcal{A}^{-1} \r) q^a q^b
\nn \\
{\rm where ~~}
\mA &=& 1 + \lp^2/\sigma_{\bm g}^2
\label{eq:key1}
\end{eqnarray}
where $\gm_{ac} \gm {}^{c b} = \delta^b_a$, $g^{ab} \equiv \l({\bm g^{-1}}\r)_{ab} $ and $q^a = g^{ab} t_b$. Note that, for $\sigma_{\bm g}^2 \gg \lp^2$, $\mathcal A \rightarrow 1$ and $\gm {}^{ab}(p;P) \rightarrow g^{ab}(p)$.

{\it I must emphasize that the above modification is non-trivial in that $\bm \gm \notin {\rm Diff}[\bm g]$ but rather describes a genuinely different spacetime -- i.e., in general, $\mathcal K [\bm \gm] \neq \mathcal K [\bm g]$, where $\mathcal K$ represents some curvature invariant. It is only when {\bf Riem}$[\bm g]=0$ that we have $\bm \gm \in {\rm Diff} [\bm g]$.}

Although a direct proof of this is difficult, involving the formidable task of computing curvature invariants for $\gm_{ab}$ in terms of those of $g_{ab}$, one can see why it must be true in at least two different ways. First, the example given below (see Sec. \ref{sec:2b}) of maximally symmetric space(time)s clearly illustrate the point, since these are the simplest space(time)s which exhibit all the symmetries of a flat space(time) but have constant, non-zero curvature. Further, Appendix \ref{app:deriv-mod-metric} gives the modified Ricci scalar, evaluated in the manner described in Sec. \ref{sec:2b}, for a deformed 2-sphere (a case which \texttt{Maple} can handle). Since a 2-sphere represents the simplest possible curved space, it very well illustrates the fact that the metrics $\gm_{ab}$ and $g_{ab}$ will generically have different curvature invariants, unless the original metric is flat. For the second argument, which is somewhat indirect, see the comment below Eq.~(\ref{eq:boxss}).

It is now straightforward to prove the following identity:
\begin{eqnarray}
\gm {}^{ab} \; \partial_a \l(\sigma_{\bm g}^2 + \lp^2\r) \partial_b \l(\sigma_{\bm g}^2 + \lp^2\r) = 4 \l(\sigma_{\bm g}^2 + \lp^2\r)
\label{eq:key2}
\end{eqnarray}
and recall that the defining equation for $\sigma_{\bm g}^2$ is the Hamilton-Jacobi equation $g^{ab} \l( \partial_a \sigma_{\bm g}^2 \r) \l( \partial_b \sigma_{\bm g}^2 \r) = 4 \sigma_{\bm g}^2$. Eq. (\ref{eq:key2}) therefore gives us our key result. It shows that $\sigma_{\bm g}^2 + \lp^2$ is indeed the geodesic distance for metric $\gm_{ab}$. That such a simple result should follow from a non-trivial modification of the metric (which, in a genuine curved spacetime, will not even have the same curvature as the original metric) is in itself remarkable. 

\underline{\it Relationship with conformal fluctuations}: We can identify $T_a=t_a/\sqrt{\mathcal A}$ as the properly normalized tangent vector w.r.t. $\gm_{ab}$. This allows us to write
\begin{eqnarray}
\gm_{ab} - \epsilon \; T_a T_b & =& \mA \; \l( g_{a b} - \epsilon \; t_a t_b \r)
\end{eqnarray}
which shows that it is the {\it induced} geometry on $\sigma_{\bm g}^2=$const surface which undergoes a conformal deformation. Extrinsic curvature of this surface can be shown to be 
\begin{eqnarray}
\overset{\star}{K}_{\mu \nu} &=& {\mathcal A}^{3/2} K_{\mu \nu} + (1/2) {\mathcal A}^{1/2} q{^k} \partial_k {\mathcal A} \; h_{\mu \nu}
\end{eqnarray}
and can be used to study $\lp$ corrections to focussing of geodesics \cite{focussing}. 

Above comments connect and contrast the present analysis with an older one based on \textit{quantum conformal fluctuations} \cite{jvn-paddy-book}, in which quantum fluctuations of the conformal factor simply lead to
\begin{eqnarray}
g^{\rm{\tiny{(q.c.f.)}}}_{ab} = \mA \; g_{ab}
\end{eqnarray}

\subsection{Green's function} \label{sec:2a}
Using identities given in Appendix \ref{app:id1}
we can show that
\begin{eqnarray}
\boxm \l( \sigma_{\bm g}^2 + \lp^2 \r) - 2D &=& \l(1 + \lp^2/\sigma_{\bm g}^2 \r) \l( \square \sigma_{\bm g}^2 - 2D \r)
\label{eq:boxss}
\end{eqnarray}
Now, in a general curved spacetime, $\sigma^2$ satisfies the equation $\square \sigma^2 = 2D + ({\rm terms~involving~curvature})$. Eq.~(\ref{eq:boxss}) therefore provides a quick confirmation of the fact that the modified spacetime is Riemann flat if (and only if) the original spacetime is Riemann flat. Using again Appendix \ref{app:id1}, we get another remarkable result
\begin{eqnarray}
\sqrt{-\gm} \;\; \boxm \l( \sigma_{\bm g}^2 + \lp^2 \r)^{-\frac{D-2}{2}} &=& \sqrt{-g} \;\; \square \l( \sigma_{\bm g}^2 \r)^{-\frac{D-2}{2}} 
\nn \\
\sqrt{-\gm} \;\; \boxm \ln \l( \sigma_{\bm g}^2 + \lp^2 \r) &\overset{(D=2)}{=}& \sqrt{-g} \;\; \square \ln \sigma_{\bm g}^2
\label{eq:leading-sing}
\end{eqnarray}
in an arbitrary curved spacetime. Since $\l( 1/\sigma_{\bm g}^2 \r)^{\frac{D-2}{2}}$ is the leading term in the scalar propagator $G(p,P | g_{ab})$ in any metric $g_{ab}$, and since the propagator satisfies 
\begin{eqnarray}
\sqrt{-g} \; \square_{\bm g}  G(p,P | g_{ab}) = \delta^{D}(p,P),
\end{eqnarray}
Eq.~(\ref{eq:leading-sing}) immediately implies that the {\it leading} modification to the propagator is obtained by replacing $\sigma_{\bm g}^2 \rightarrow \sigma_{\bm g}^2 + \lp^2$, which generalizes the flat spacetime result for which, of course, this replacement gives {\it exact} propagator.

\subsection{Spacetime singularities} \label{sec:2b}
We now turn to the second relevant implication of minimal length: the modification to invariant properties of spacetime such as it's curvature invariants. Once the world function is known in a given spacetime, a symbolic package such as \texttt{Maple} can be used to obtain expressions for, say, any modified curvature invariant $\overset{\star}{\mathcal{K}}(p;P,\lp)$. It's behaviour at any spacetime event $P$ can then be deduced from $\overset{\star}{\mathcal{K}}(P,\lp) = \underset{p \rightarrow P}{\lim} \overset{\star}{\mathcal{K}}(p;P,\lp)$. Unfortunately, exact expression for the world function is rarely obtainable, and it's approximate expansion in coordinate intervals (which is what is often resorted to in the literature) is often inadequate near singularities. However, the following results can be obtained in a straightforward manner:

{\it 1. Flat spacetime}: 
By construction, for a flat spacetime, $\overset{\star}{R}_{abcd}(p;P)=0$, and hence 
$\overset{\star}{R}_{abcd}(P,\lp) = \underset{p \rightarrow P}{\lim} \overset{\star}{R}_{abcd}(p;P,\lp)=0$.

{\it 2. Symmetric spacetimes}:
For maximally symmetric spaces, $\Omega(p,P)$ is known in a closed form, and a \texttt{Maple} computation gives: $\overset{\star}{\mathcal{K}}(P,\lp) = {\mathcal{K}}(P) \l[ 1 + c_{\mathcal{K}} {\mathcal{K}}(P) \lp^2 \r]$, where $c_{\mathcal{K}}$ is a numerical coefficient. 

{\it 3. Schwarzschild singularity}: 
World function for this case is not known in closed form, but {\it if} one restricts to radial timelike geodesics near the singularity ($r=0$), one finds that the modified Kretschmann scalar behaves as $1/(1+\epsilon)\lp^4$ instead of $M^2/r^6$. Although still divergent since $\epsilon=-1$, the divergence is of a very different character and, being $M$ independent, can presumably be regularized. 

For non-singular spacetime events, it should be possible to employ Riemann Normal Coordinates (RNC) and establish {\it (1)} and {\it (2)} generically. Investigation is currently in progress as to whether something generic can be said about {\it curvature singularities}.

\section{Discussion} \label{sec:discussion}
The modified metric derived here has several conceptual connotations. I begin with the most relevant one - the analytic structure of the metric in the coincidence limit (this has also been emphasized by Brown \cite{brown-qg, brown-qg2}): the term $(1 + \mQ)^{-1}$, with $\mQ=\lp^2/\sigma^2$ in the modified metric can be represented as a series $\mathcal S = \sum_{n=0}^{\infty} (- \mQ)^n$ when $|\mQ| < 1$, i.e., when $\sigma^2 > \lp^2$. It is quite possible that one arrives at some such series by doing a perturbative analysis for small $\mQ$. Any such series would not be valid beyond it's region of convergence ($|\mQ|=1$), and there might be many different ways to extract from it a term 
non-analytic at $\sigma^2=0$. In the present analysis, the term $(1 + \mQ)^{-1}$ which arises naturally is the analytic continuation of the series $\mathcal S$ beyond it's region of convergence. (See also \cite{brown-qg}.)

As already stated, when spacetime is flat, the modified metric can be mapped back, via a {\it singular} coordinate transformation, to the original flat spacetime, but with a region (a $``$hyperboloid") of size $\lp$ around the base point ($X$ above) removed. (Incidentally, this also naturally suggests existence of a {\it maximal} acceleration, $a_{\rm max} \equiv \lp^{-1}$ in flat spacetime. See Appendix \ref{app:max-acc}.) 
Usually, for a smooth curved manifold with metric $g_{ab}$, one employs the standard metric expansion in RNC [$y^i = x^i-X^i$]
\begin{eqnarray}
g_{a b}(x;X) \approx \eta_{a b}  - ({1}/{3}) R_{a c b d}(X)~y^c y^d + O(y^3) 
\label{eq:rnc}
\end{eqnarray}
However, there are (at least) two important cases when such an approximation is unjustified: (a) for extremely small spacetime intervals, for which continuous structures such as $R_{a b c d}$ are unlikely to make much sense, and (b) near spacetime singularities, where the curvature tensor and/or it's derivatives might blow up. Motivated by several earlier results on minimal length, we have arrived at a metric which can serve as an effective metric at small scales. The non-locality of this metric is best understood by comparing it with the above RNC expansion: in the modified case, the metric at $x \in \mathcal N(X)$ depends not only on curvature at $X$, but also on the spacetime interval between $x$ and $X$. The resultant non-locality is therefore natural, an outcome of our starting assumption that spacetime intervals have a lower bound. This also suggests analysing the modified metric in the context of ``spacetime foam". (See also \cite{garay-foam}.)

It would be insightful to set up an \textit{effective action} which is extremized by the modified metric. Such an action will necessarily be \textit{non-local}, but should be relatively straightforward to set up given the disformal form of the metric. One earlier work with somewhat similar motivation is by DeWitt (see the second reference in \cite{dewitt-1964}). 

As already emphasized, our result captures the essence of several earlier works on the topic. More importantly, the closed form expression for the modified metric should make it possible to perform explicit computations in the modified geometry. All one needs is an expression for the world function $\Omega(p,P)$ in the background geometry, either under some (valid) approximation, or exact if available. In fact, our analysis also suggests a conceptually appealing possibility in which the world function might play a more fundamental role than the spacetime metric itself. This is very much possible, since almost all the information about spacetime geometry can be shown to be encoded in the coincidence limit (denoted below by ``$[\ldots]$") of covariant derivatives of $\Omega(p,P)$. For e.g. \cite{poisson-lrr}, 
\begin{eqnarray}
g_{a'b'} = g_{ab} &=& \l[{\nabla}_a {\nabla}_b \Omega(x,x') \r] = \l[{\nabla}_{a'} {\nabla}_{b'} \Omega(x,x') \r] 
\nn \\
R_{a' (c' d') b'} &=& \l(3/2\r) \l[{\nabla}_a {\nabla}_b \nabla_{c} \nabla_{d} \Omega(x,x') \r]
 \nn
\end{eqnarray}
and so on. One can therefore ``re-construct" spacetime from $\Omega(p,P)$, and study how modifying $\Omega(p,P)$ affects the re-constructed spacetime, which is essentially what we have done. (This might have overlap with some interesting recent ideas, mainly due to Achim Kempf \cite{kempf}.) In this sense, the ``local" nature of familiar gravitational actions also seems to be an illusion which would break down when $2 \Omega(p,P) \lesssim \lp^2$. 

Finally, I must mention that the entire analysis presented here is for timelike/spacelike separated events. The case of null separated events bring in several subtleties, both technical, and more importantly conceptual, since all arguments in the literature for minimal length deal with bounds on timelike/spacelike intervals only, and there are no such arguments for null separated events. We hope to study this case in subsequent work.

{\it Acknowledgements --} I thank T. Padmanabhan, L. Sriramkumar, and J. Narlikar for useful comments and suggestions, and also A. Kempf for pointing out the second reference in \cite{kempf}. Part of this research was supported by post-doctoral fellowship from NSERC and AARMS, Canada. I also thank IUCAA, Pune and CTP Jamia, New Delhi for hospitality during completion of this work. All symbolic computations relevant for this work have been done in \texttt{Maple} using GRTensorII \cite{grtensor}.
\appendix 
\section{Derivation of Eq. (4)} \label{app:deriv-mod-metric}
We look for a re-parametrization of flat space(time) such that geodesic distances have a natural lower bound, and we do so by introducing a ``hole" in flat space(time). The analysis is most easily done in flat Euclidean space, and then analytically continuing back to Lorentzian signature. Consider therefore an event $\mathcal P$ in a $D$ dimensional flat Euclidean space (for simplicity, I set it's coordinates $X^k=0$ without loss of generality; they can be easily restored in final expressions). The result in Eq. (4) is derived via following steps: start with standard Cartesian coordinates for flat space, transform to spherical-polar like coordinates with the ``radial" coordinate being the geodesic distance $\sigma_E$, introduce a new coordinate via $\sigma_E \rightarrow \sqrt{\sigma_E^2 + \lp^2}$, and then revert to Cartesian like coordinates using standard transformations. (For the Lorentzian case, the only difference is the appearance of hyperbolic, instead of standard trig, function in one of the transformations.) These steps are straightforward, and lead to the following re-parametrization 
\begin{eqnarray}
x_E^k \rightarrow \sqrt{ 1 - \frac{\lp^2}{\delta_{ab} {x}_E^a {x}_E^b} } \; {x}_E^k
\end{eqnarray}
which must be treated as an active diffeomorphism (notice that it becomes singular in the coincidence limit). The above diffeomorphism then yields 
the Euclidean analogue of the metric $\overset{\star}{\eta}_{a b}$ given in Eq.~(\ref{eq:dm-metric}) of the paper (in which we have re-introduced the coordinates $X^k$ of $\mathcal P$). After some further manipulations, we can also obtain the Euclidean analogue of the form given in Eq.~(\ref{eq:key1}):
\begin{eqnarray}
\overset{\star}{\delta}_{ab} &=& A \delta_{ab} - \l( A - A^{-1} \r) t^E_{a} t^E_{b}
\label{eq:app-euclidean}
\end{eqnarray}
where $t^E_a=\delta_{a b} x_E^b/\sqrt{\delta_{mn} x_E^m x_E^n}$ with $t_E^2 = + 1$ is the normalized geodesic tangent vector connecting $X^k$ to $x_E^k$.

{\it Analytic continuation to Lorentzian signature}: The final step is the analytic continuation from Euclidean to Lorentzian signature, in which the only non-triviality concerns the $t^E_{a} t^E_{b}$ term in the metric. For the Lorentzian signature, the normalized geodesic tangent vector is given by
\begin{eqnarray}
t_a &=& \frac{\eta_{a b} x^b}{\sqrt{\epsilon \; \eta_{mn} x^m x^n }}
\nn \\
t^2 &=& \epsilon = \pm 1
\end{eqnarray}
The correct analytic continuation, $t^E_{a} t^E_{b} \rightarrow \epsilon t_{a} t_{b}$, is most easily deduced by starting from 
\begin{eqnarray}
\overset{\star}{\eta}_{a b}(x; X) = A \eta_{ab} - K \epsilon \l( A - A^{-1} \r) t_{a} t_{b}
\label{eq:app-minkowski-1}
\end{eqnarray}
and showing that the desired result requires $K=1$. More specifically, evaluate the inverse metric 
\begin{eqnarray}
\overset{\star}{\eta} {}^{a b}(x; X) &=& A^{-1} \eta^{ab} + F \epsilon \l( A - A^{-1} \r) q^{a} q^{b}
\\
F &=& \l[{1 + (K^{-1}-1) A^2}\r]^{-1} \; ; \;\; q^a = \eta^{ab} t_b
\nn 
\end{eqnarray}
To fix $F$, we appeal to the Hamilton-Jacobi equation, which yields
\begin{eqnarray}
\overset{\star}{\eta} {}^{a b} \partial_a (\sigma^2 + \lp^2) \partial_b (\sigma^2 + \lp^2)
=
4 \sigma^2 \l[ F A + \frac{1-F}{A} \r]
\end{eqnarray}
To get the desired result $4 (\sigma^2 + \lp^2) = 4 A \sigma^2$ on the RHS, we must therefore have $F=1$, which fixes $K=1$. 

It is instructive to explicitly analyse what our modification gives when applied to a 2-sphere, the simplest possible example of a curved space. The geodesic distance between points $p$ and $P$ on a 2 sphere of radius $L$ is given by 
\begin{eqnarray}
\sigma &=& L \cos^{-1} \Theta
\nn \\
\Theta &=& \cos \theta \cos \theta' + \sin \theta \sin \theta' \cos \l(\phi-\phi'\r) 
\end{eqnarray}
An explicit \texttt{Maple} computation gives the Ricci ``bi"-scalar for the modified metric as
\begin{eqnarray}
\overset{\star}{R}(p;P,\lp) &=& \frac{2}{L^2} \l[ 1 + \frac{\lp^2 \; f(\sigma/L)}{L^2} \r]
\end{eqnarray}
where 
\begin{eqnarray}
f(x) &=& \frac{x^2 + 3 \l( x \cot x - 1 \r)}{x^4}
\nn \\
\lim \limits_{x \rightarrow 0} f(x) &=& -\frac{1}{15}
\\
\end{eqnarray}
Using this, we finally obtain
\begin{eqnarray}
\overset{\star}{R}(P,\lp) &=& \underset{p \rightarrow P}{\lim} \overset{\star}{R}(p;P,\lp)
\nn \\
&=& \frac{2}{L^2} \l( 1 - \frac{1}{15} \frac{\lp^2}{L^2} \r)
\end{eqnarray}

\section{Some Identities} \label{app:id1}
The following identities, which are relevant for the analysis of Green's function in a $D$ dimensional curved spacetime, can be proved in a straightforward (although lengthy) manner using the modified metric. To begin with, one may use the {\it matrix determinant lemma}: ${\mathrm{det}} \l(\rm \bf M + \rm \bf u \rm \bf v^{\mathrm{T}}\r) = \l(\mathrm{det} \; \rm \bf M\r) \times \l( 1 + \rm \bf v^{\mathrm{T}} \rm \bf M^{-1} \rm \bf u \r)$, where $\rm \bf M$ is an invertible square matrix, and $\rm \bf u, \rm \bf v$ are column vectors (of same dimension as $\rm \bf M$), to obtain: $\sqrt{-\gm} = \sqrt{-\gdummy } \l( 1 + \lp^2/\sigma_{\bm g}^2 \r)^{(D-2)/2}$. Using this, one can show that (with $\mQ \equiv \lp^2/\sigma^2$)
\begin{eqnarray}
\boxm \l( \sigma_{\bm g}^2 + \lp^2 \r)^{-\frac{m}{2}} &=& \l( 1 + \Psi \r)^{-\frac{m}{2}} \Biggl[ \square \l( \sigma_{\bm g}^2 \r)^{-\frac{m}{2}} 
\nn \\
&-&
 m (m-D+2) \l( \frac{\Psi}{1 + \Psi} \r) \l( \sigma_{\bm g}^2 \r)^{-1-\frac{m}{2}}
\Biggl]
\nn \\
\boxm \ln \l( \sigma_{\bm g}^2 + \lp^2 \r) &=& \square \ln \sigma_{\bm g}^2 + 
\l( \frac{\Psi}{1 + \Psi} \r) \l(\frac{4 - 2D}{\sigma_{\bm g}^2}\r)
\nn
\end{eqnarray}
\section{Minimal length and Maximal acceleration} \label{app:max-acc}

That our Lorentz invariant modification of geodesic distances $\sigma^2 \rightarrow \sigma^2 + \lp^2$ has implications for maximal acceleration in flat spacetime is essentially a consequence of the fact that uniformly accelerated observers with acceleration $g$ in flat spacetime are given by contours of $\sigma^2(p,P)=g^{-2}$ where $P$ is the origin (the points on the contours are therefore connected to $P$ by spacelike geodesics). A lower bound on $\sigma^2$ should therefore also imply an upper bound on acceleration. One can explicitly demonstrate this in flat spacetime as follows. For simplicity, we will work in $(1+1) D$ which suffices for our purpose. Set up the deformed metric in a spacelike neighbourhood of $P$, in the so called right Rindler wedge, in Rindler coordinates. The standard Rindler metric is given by $\DM s^2 = - N(x)^2 \DM t^2 + \DM x^2$, where $N(x)=1+gx$ and $x \in [-1/g, \infty)$. The acceleration of any $x=$const. curve is given by $a=g/(1+gx)$, which {\it diverges} at $x=-1/g$, the Rindler horizon. To construct the deformed metric, we need the geodesic distance between two events in Rindler coordinates; this is given by $\sigma^2(p,P) = g^{-2} \l[ N(x)^2 + N(X)^2 - 2 N(x) N(X) \cosh g \Delta t \r]$. Although the final form of the metric is unwieldy, one may set up a \texttt{Maple} routine to evaluate the acceleration of a $x=$const. curve;  this yields $\overset{*}{a}=g/\sqrt{N(x)^2+g^2 \lp^2}$. Since $N(x)=0$ at $x=-1/g$, we get $\overset{*}{a}_{\rm max} = 1/\lp$.

Note that this result hinges on the fact that uniformly accelerated observers in flat spacetime move along orbits of the boost Killing field, which are hyperbola's given by contours of the geodesic interval $\sigma^2$. It's generalization to curved spacetime is currently under investigation.


\end{document}